\documentclass[lettersize,journal]{IEEEtran}
\usepackage{amsmath,amsfonts}
\usepackage{algorithmic}
\usepackage{algorithm}
\usepackage{array}
\usepackage[caption=false,font=normalsize,labelfont=sf,textfont=sf]{subfig}
\usepackage{textcomp}
\usepackage{stfloats}
\usepackage{url}
\usepackage{verbatim}
\usepackage{graphicx}
\usepackage{cite}
\usepackage{accents}           %
\usepackage{hyperref}
\usepackage{booktabs}
\usepackage{multirow}
\usepackage{bm}             

\DeclareMathOperator*{\argmin}{arg\,min}
\newcommand{\norm}[1]{\left\lVert#1\right\rVert}
\newcommand{\figref}[1]{Figure \ref{#1}}
\newcommand{\secref}[1]{Section \ref{#1}}
\usepackage{array}
\usepackage{bbold}

\usepackage{pdfpages}       
\newboolean{showcomments} 
\setboolean{showcomments}{false}
\newcommand{\slow}[1]{\ifthenelse{\boolean{showcomments}}
{ \textcolor{red}{(SL: #1)}}{}}
\newcommand{\ldw}[1]{\ifthenelse{\boolean{showcomments}}
{ \textcolor{blue}{(LW: #1)}}{}}
\newcommand{\yx}[1]{\ifthenelse{\boolean{showcomments}}
{ \textcolor{green}{(YX: #1)}}{}}
\newcommand{\hx}[1]{\ifthenelse{\boolean{showcomments}}
{ \textcolor{brown}{(HX: #1)}}{}}

\begin{document}

\title{Synchrophasors and Synchrowaveforms for the Distribution Grid: The SoCal 28-Bus Dataset}
\author{
    Yiheng Xie, 
    Lucien Werner, 
    Kaibo Chen, 
    Thuy-Linh Le, 
    Christine Ortega, 
    Steven Low
\thanks{The authors are with the Department of Computing and Mathematical
Sciences, California Institute of Technology, Pasadena, CA 91125 USA\\
(e-mails: \{yxie5,lwerner,chenkb,thuylinh,cortega,slow\}@caltech.edu).}
\thanks{The authors are supported by the Resnick Sustainability Institute and S2I.}
\thanks{
}
}


\IEEEpubid{}

\maketitle
\begin{abstract}
This paper presents an open-access dataset of phasor and waveform measurements from a real-world operational distribution network. The dataset encompasses diverse generation resources (including solar panels, fuel cells, natural gas generators, and utility interconnections), loads (including large-scale electric vehicle charging, data centers, central cooling, and offices), as well as topology changes (such as line outages and load transfers). Coverage is dense: all buses with non-zero power injections are instrumented. The availability of dense synchrophasor measurements enables a range of applications, including state estimation, system identification, power flow optimization, and feedback control. Additionally, the dataset includes synchronized waveforms, which enable the analysis of harmonics, sub-cycle events, dynamic grid impedance, and fast-timescale stability. Finally, the dataset provides time-varying circuit topology and parameters. Data collection began in 2023, and new data is generated continuously and made available online.
\end{abstract}


\section{Introduction}
Distribution grids today are experiencing rapid transformation with the proliferation of distributed energy resources (DERs), which stress the system in both steady-state operations, transient dynamics, and future planning. Issues such as stability, uncertainty, and constraint violations have been widely reported in distribution networks \cite{petinrin_impact_2016, haque_review_2016, khani_impacts_2012, rafique_bibliographic_2022, fachrizal_combined_2021, wen_analysis_2016}. 
Solving these challenges requires high-quality data, including both the circuit topology and time-series measurements. 
A comparison with existing metering and available datasets reveals that the currently available data is insufficient for detailed analysis such as computing the phasor-based network power flow. 


\begin{figure*}[!ht]
\centering
\includegraphics{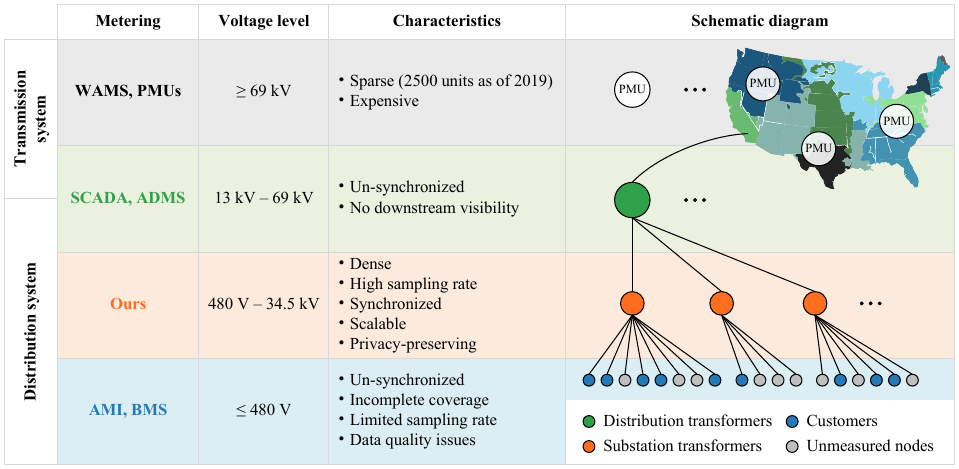}
\caption{A summary of power systems metering approaches. 
WAMS: Wide Area Measurement System. PMUs: Phasor Measurement Units. SCADA: Supervisory Control and Data Acquisition. ADMS: Advanced Distribution Management Systems. AMI: Advanced Metering Infrastructure. BMS: Building Management Systems. The map is adapted from \cite{ahmad_guardians_2024}.
}
\label{fig:approach}
\end{figure*}

\subsection{Limitations in Legacy Systems}
The common sources of distribution grid data are supervisory control and data acquisition (SCADA) systems in industrial settings \cite{thomas_power_2017}, building management systems (BMS) in campus settings, and advanced metering infrastructure (AMI) in residential settings \cite{rashed_mohassel_survey_2014}. However, there are several important limitations. 
\subsubsection{Sparsity} Traditional PMU deployment is limited to sparse locations in the transmission grid due to their prohibitive installation cost \cite{us_department_of_energy_big_nodate, noauthor_estimated_2021, dusabimana_survey_2020}. In distribution systems, AMI meters are installed at individual customer service points. However, AMI deployment and data availability have historically been limited by installation cost, data management challenges, and the large number of customers, although coverage continues to expand.
\subsubsection{Sampling rate and latency} The sampling intervals of existing systems are constrained by data bandwidth and storage capacity, and are typically 1 to 15 minutes in the distribution grid \cite{thomas_power_2017, rashed_mohassel_survey_2014}. Monitoring and closed-loop control further require real-time data transfer. Many legacy metering systems are not designed for these applications.
\subsubsection{Synchronization} Although modern switchgears and inverters are capable of capturing waveforms and phasors, these instruments are rarely synchronized to a common time reference, and the \textit{relative} phase angle between two different nodes in the network cannot be obtained \cite{dusabimana_survey_2020}. AC power flow formulations require phase angles as part of the system of equations. 
\subsubsection{Practical issues} System-level analysis requires a consistent data format, yet data harmonization across heterogeneous hardware, installed over several decades, is a significant undertaking.
Finally, from our experience collaborating with utilities and campus facilities departments, many existing data collection systems are often plagued with missing data, mis-calibration, mis-labeling, and other quality issues \cite{della_giustina_electrical_2014}.

These limitations are largely due to the fact that traditionally, distribution grids mainly consist of passive circuit elements over which system operators exert very limited control. As such, most existing distribution grids are designed and operated with limited visibility, with scarce time-series measurements \cite{della_giustina_electrical_2014} and a lack of accurate topology information \cite{weng_distributed_2017}. With the proliferation of DERs, modern distribution systems are increasingly operated towards the limit while the availability of controllable devices also increases. Legacy metering systems are no longer sufficient to support modern distribution grid operations and planning.

A summary of existing metering approaches against ours is provided in \figref{fig:approach}. In comparison, our approach is different in that the meters are densely deployed unlike typical SCADA, ADMS, and transmission system PMUs; offer a higher sampling rate than typical AMI and BMS systems; are synchronized with a global phase reference unlike most distribution grid meters; and preserve privacy by aggregating individual customers.

\subsection{Related Datasets}
There are very few real-world synchrophasor datasets that are publicly available today. Traditionally, due to their prohibitive cost, PMUs are deployed sparsely in transmission systems \cite{banning_foa_2021}. The work most closely related to ours is the Open $\mu$PMU project jointly led by Lawrence Berkeley National Laboratory (LBNL) and the California Institute for Energy and Environment (CIEE) \cite{stewart_phasor_2016, stewart_open_2016}. To the best of our knowledge, this is the only open-access synchrophasor dataset from a real-world system. The open-access portion contains measurements of 3 nodes over a 3-month period in 2015, collected on the LBNL campus. Our dataset is an order of magnitude larger in both the number of nodes metered and the duration of data available, and additionally includes synchrowaveforms.

Another related work is the FNET/GridEye project \cite{zhu_fnetgrideye_2020, liu_distribution_2017, liu_recent_2016}, which has over 250 frequency disturbance recorders deployed across the continental U.S. (and some internationally). Our approach is similar to theirs in that a network of sensors are deployed and synchronized across long distances via GPS. However, the recorders lack the ability to measure current. 

While synchrophasors provide a description of steady-state single-frequency sinusoidal signals, synchrowaveforms (also known as point-on-wave data) allow sub-cycle description of distortions and transient events. 
Recently, the use of synchrowaveform data has started to gain traction \cite{mohsenian-rad_synchro-waveforms_2023, hamed_mohsenian-rad_synchro-waveform_nodate}. 
With growing penetration of inverter-based resources (IBRs), the transient dynamics and stability become more relevant. Furthermore, in distribution systems, the steady-state assumptions underlying phasors are often violated. Large distortions in waveform signals are often observed due to nonlinear connected devices. Synchrowaveforms are therefore increasingly relevant for future distribution grids.

Nonetheless, open-access synchrowaveform data is scarce at the time of writing. The Grid Event Signature Library (GESL) offers an open-access phasor and waveform dataset, collected from 10 utilities in the U.S. primarily for the purpose of event detection and analysis \cite{GESL2023Dataset_ORNL_LLNL}. The data is generated from sparse and heterogeneous sources across a vast area.
The EPFL-campus dataset \cite{romano_enhanced_2014} provides point-on-wave measurements at 50 kHz from a 20 kV campus substation, captured during step changes in a battery energy storage system. The dataset is generated by one meter at a single busbar.
In contrast, our meters are densely deployed in one distribution circuit so that Kirchhoff's laws and Ohm's law apply. 

As a result of the lack of real-world data, modern distribution grid research is often tested on simulated circuits such as the IEEE n-bus systems. There is a lack of real-world test cases for which the circuit parameters and time-series originate from a physical system. This dataset can be used to build a testbed based on a real-world distribution circuit. 
See Table \ref{tab:related_datasets} for a summary of related datasets.

\begin{table*}[ht]
\centering
\caption{Comparison of related public-access power system and metering datasets.}
\label{tab:related_datasets}
\begin{tabular}{lccccc}
\toprule
\textbf{Dataset} &
\textbf{Observability} &
\textbf{Number of} &
\textbf{Measured} &
\textbf{Topology \&} &
\textbf{Available Data} \\
&
\textbf{(System-wide)} &
\textbf{Measured Nodes} &
\textbf{Quantities} &
\textbf{Parameters} &
\textbf{Time Span} \\
\midrule
FNET/GridEye &
No &
$>$300 &
Voltage, frequency, angle &
Partial &
$>$ 20 years \\
\cite{liu_distribution_2017, liu_recent_2016, zhu_fnetgrideye_2020} &
&
&
(phasors) &
&
 \\
\midrule
Open $\mu$PMU &
Unknown &
3 &
Voltage and current &
Yes &
3 months \\
\cite{stewart_open_2016, stewart_phasor_2016, dusabimana_survey_2020} &
&
(public portion) &
(phasors) &
&
(public portion) \\
\midrule
GESL &
No &
$>$2{,}600 events &
Voltage and current &
No &
N/A \\
\cite{GESL2023Dataset_ORNL_LLNL, us_department_of_energy_big_nodate} &
&
&
(phasors \& waveforms) &
&
(event-based) \\
\midrule
AMI datasets &
Typically no &
Varies &
Energy, voltage magnitude, &
Typically no &
Varies \\
\cite{pecanstreet_dataport,cer_smart_meter_2012,uk_smart_meter_london,smart_star_dataset,refit_dataset_2015} &
 &
 &
power factor (some) &
 &
 \\
\bottomrule
\end{tabular}
\end{table*}
\subsection{Contributions}
To the best of our knowledge, this is the first densely collected synchrophasor and synchrowaveform dataset in a distribution system -- there are no open-access datasets that simultaneously meet the following requirements: high spatial resolution (dense deployment) enabling system observability, high temporal resolution, microsecond-level synchronization accuracy for accurate phase reference, and low latency for real-time applications. Additionally, these objectives are achieved at a cost that is suitably low for the economic scale of medium- and low-voltage grids. 

The paper is organized as follows. Section~\ref{sec:dataset} presents the dataset. Section~\ref{sec:methods} describes the metering system implementation. Section~\ref{sec:errors} characterizes data accuracy. Section~\ref{sec:applications} presents example applications enabled by this work. Section~\ref{sec:conclusion} concludes the paper.
\section{Dataset}
\label{sec:dataset}
In this section, we describe three types of data made available: synchrowaveforms, synchrophasors, and circuit topology and parameters (digitized single-line diagrams). We provide the most granular and unprocessed data possible so as to enable a wide range of use cases. 
A collection of helper functions is provided to process the data into common useful formats. \textbf{The full dataset, documentation, and code are accessible: }\url{https://github.com/caltech-netlab/digital-twin-dataset}.\footnote{DOI: \href{https://dx.doi.org/10.21227/hsfq-fq78}{10.21227/hsfq-fq78}}

\begin{figure}[b]
    \centering
    \includegraphics[width=1.0\linewidth]{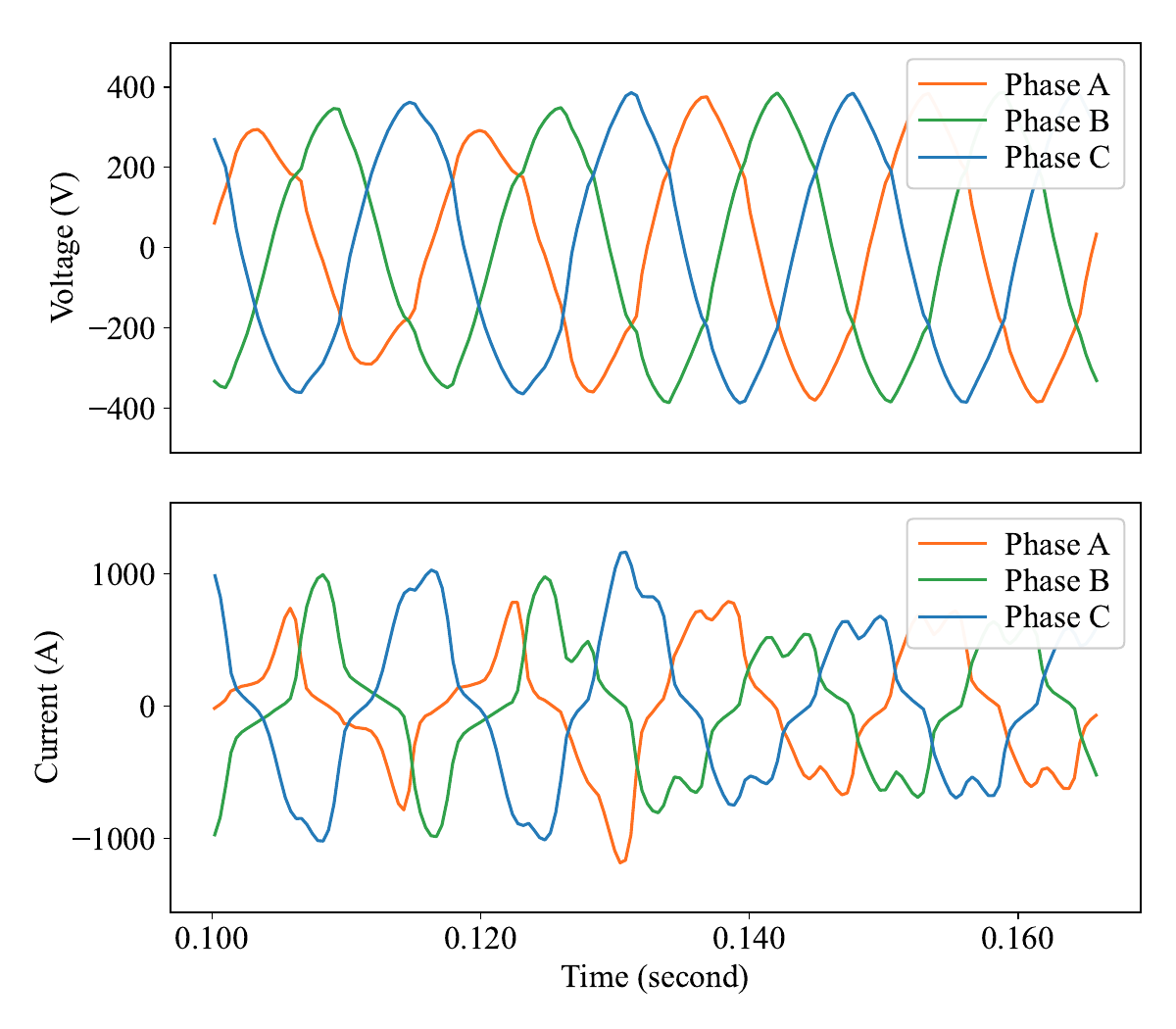}
    \caption{Three-phase voltage and current injection waveform measurement at bus 1118. 
    Phase imbalance, harmonic distortion, non-unitary power factor, and a transient event are present.}
    \label{fig:data_waveform}
\end{figure}

\paragraph{Dataset at-a-glance.} The physical size of the circuit is on the order of 1,000 ft. The circuit consists of 3-phase underground untransposed lines and 3-phase Delta-Wye transformers connecting 2 (normally disconnected) radial distribution circuits, where the leaf nodes are the load buses. The lines have R/X ratios on the order of 0.1 to 1.0. Synchrowaveforms are recorded at 2.5 kHz. Synchrophasors are derived from the waveforms. The voltage total harmonic distortion (THD) is typically 1\% to 2\% while the current THD varies greatly depending on the load. The voltage phase imbalance is typically less than 0.1\% and the current phase imbalance typically varies from 0.1\% to 30\% depending on the load. The majority of the loads are offices and research facilities unless otherwise indicated. Data is available from 2023 to the time of writing.

\subsection{Synchrowaveforms}
As described in \cite{mohsenian-rad_synchro-waveforms_2023}, synchrowaveforms are a set of time-synchronized current and voltage waveforms at multiple locations in a power system. This is the most authentic representation of signals in power systems. We simultaneously capture voltage and current waveforms across all sensors. The measurement timestamps across different sensors are synchronized as described in Section \ref{sec:methods; subsec:sync}. 
The waveform sampling rate is 2.5 kHz. Due to hardware limitations, our meters do not support continuous recording. Each waveform is 1 second in length and a waveform is captured every 10 seconds, i.e. 10\% temporal coverage. As such, the majority of the data is best-suited for steady state analysis. In addition, rare transient events are also captured in some cases. An example is shown in \figref{fig:data_waveform}.

\subsection{Synchrophasors}

\begin{figure}[b]
\centering
\includegraphics[width=1.0\linewidth]{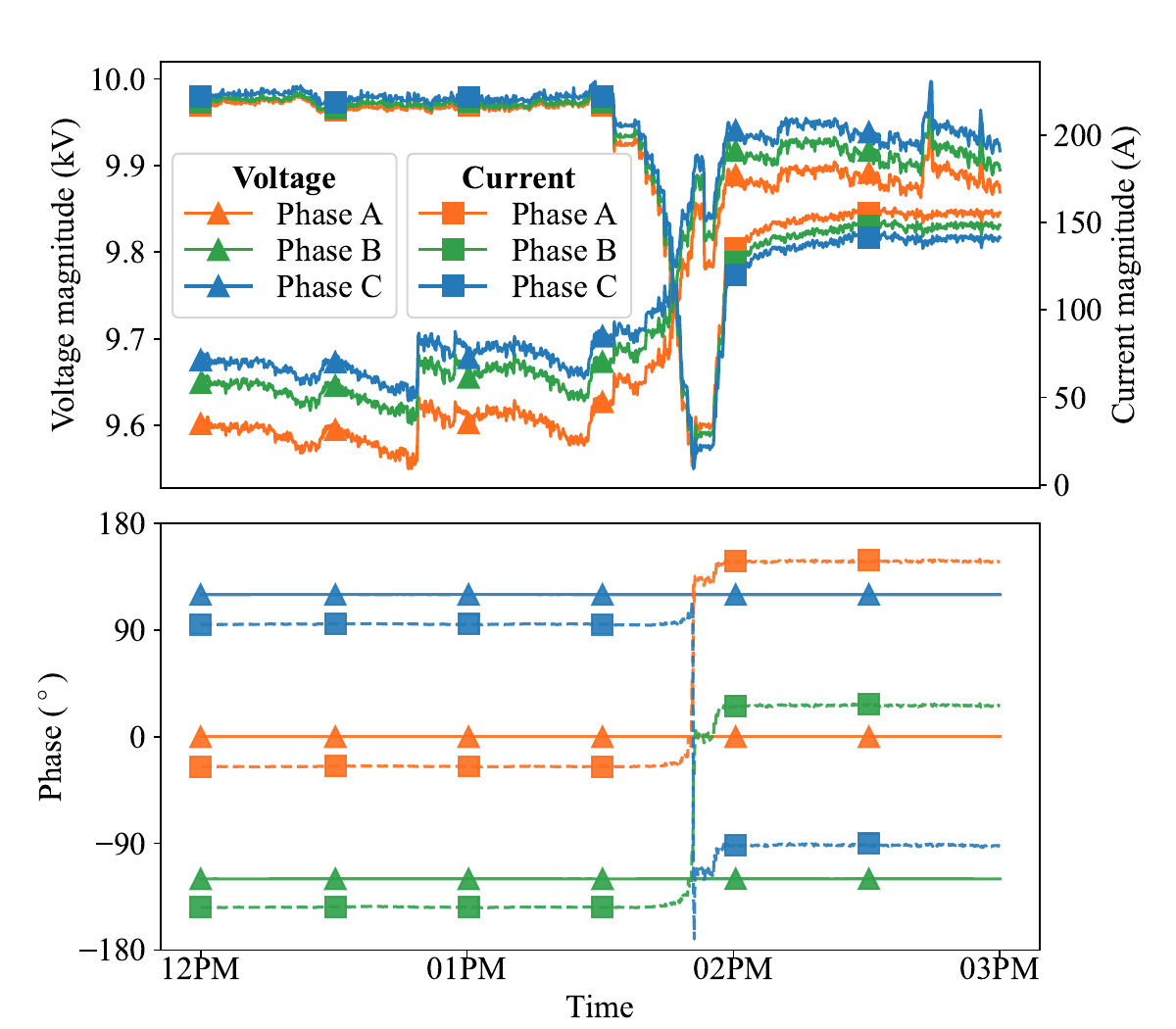}
\caption{Three-phase voltage and current injection phasor measurement (magnitude and phase) at bus 1033. Current polarity reversal is observed due to a nearby generator coming online.}
\label{fig:data_phasor}
\end{figure}

We compute the voltage and current synchrophasors from synchrowaveforms. The data columns are time, phase angle, and magnitude. One phasor at the fundamental component is extracted from each one-second waveform via fast Fourier transform (FFT). Since we release the raw waveform data, interested users can calculate additional harmonic phasors, adopt a more sophisticated frequency estimation method, and generate more than one phasor from each waveform. Due to the sampling limitations by the low-cost hardware, the synchrophasor measurements are not guaranteed to satisfy accuracy standards such as IEEE C37.118.1-2011. An example synchrophasor timeseries is shown in \figref{fig:data_phasor}.


\begin{figure*}
\centering
\includegraphics{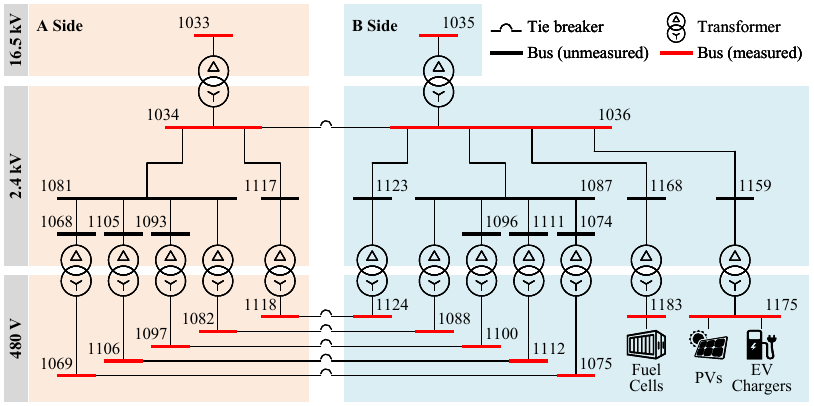}
\caption{A sub-circuit of this dataset. Zero-impedance elements are removed in plotting. Tie breakers are open during normal operating conditions.}
\label{fig:data_topology}
\end{figure*}

\subsection{Circuit Topology and Parameters}

For the distribution circuit metered, we provide the circuit topology and parameters, which we refer to as digitized single-line diagrams. The data contains information such as line and bus connectivity, conductor material and geometry, transformer nameplate ratings, and time-varying switch and breaker status. 

The series and shunt resistance, capacitance, and inductance for distribution lines can be determined from the digitized single-line diagrams using Carson's equations \cite{kersting_distribution_2012}. The transformer series impedance, turns ratio and tap positions are also provided. Together, the network admittance matrix can be calculated. Example system models in time and phasor domains are formulated in \secref{sec:applications}.

It is important to note that the digitized single-line diagrams contain zero-impedance elements (e.g. short lines and closed switches) as well as infinite-impedance elements (e.g. open switches). This is the \textit{physical asset network}, from which the \textit{electrical network} can be obtained by removing edges corresponding to zero- and infinite-impedance elements and combining buses connected by zero-impedance elements. \figref{fig:data_topology} shows the electrical network of the metered circuit.


\section{Implementation}
\label{sec:methods}
This section describes the methods for implementing our sensor network. We clarify that the focus is not in the meter hardware itself, but the approach in using readily available commercial meters to obtain high-value data in a scalable way.

\subsection{Siting}
We measure the bus voltage and the current on all incident lines for all buses with non-zero current injection. In a typical radial distribution network, this corresponds to measuring the substation transformer as the root node, and choosing a voltage level (e.g. 480 V) as the leaf nodes. The subtree connected to each leaf node is replaced by one injection. \figref{fig:data_topology} shows an example of this siting approach. In our case, all non-linear circuit elements (e.g. time-varying loads and generators) are measured behind the meter such that the remaining circuit consists of linear circuit elements only (e.g. lines, switches, and transformers).

Optimal PMU placement is a well-studied problem in transmission systems \cite{manousakis_taxonomy_2012}. In general, the goal is to deploy the minimum number of PMUs to achieve system observability. A system is \textit{observable} if the system state can be determined from measurements and a system is \textit{numerically observable} if the measurement Jacobian is full-rank \cite{sodhi_optimal_2010}. 
The observability conditions vary for different system models and problem formulations, and hence are specific to the particular application scenario. Our dataset is numerically observable for the purpose of phasor-domain state estimation described in \secref{sec:state_estimation}.

\subsection{Installation}

\begin{figure}[ht!]
\centering
\includegraphics[width=1.0\linewidth]{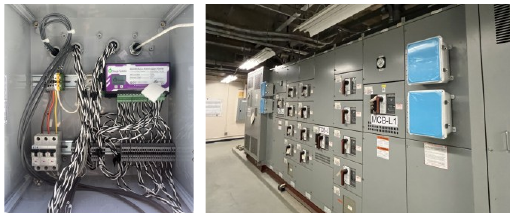}
\caption{Left: a meter box including voltage connections (3-phase 4-wire), current transformer connections (twisted pairs), and an Ethernet connection. Some sites may additionally include an edge computing device, a wireless communication module, and a battery power backup system. Right: a typical installation site with transformers and switchgear.}
\label{fig:installation}
\end{figure}

Most meters are installed on the secondary side of distribution transformers, typically at the 277/480 V level. Installation at low voltage levels significantly reduces the cost in both material and labor. In most cases, the installation does not require de-energizing the switchgear, which minimizes the disruption to customers and reduces scheduling difficulty. An example meter box and installation site is shown in \figref{fig:installation}.

Additionally, at least one meter is installed at the substation transformer at the 2.4 kV or 16.5 kV level to measure the root node. In this scenario, legacy metering equipment often exists, whose existing potential and current transformers can be used so as to avoid accessing the live transformer switchgear. We provide a list of material and cost estimate in Table \ref{tab:material_cost}.

\subsection{Time Synchronization}
\label{sec:methods; subsec:sync}
A key contribution in our work is the high-accuracy time synchronization across affordable electrical meters. Since AC power flow can be driven by small differences in the voltage phase angles, synchronization error must be within a fraction of a degree of a 50- or 60-Hz cycle \cite{stewart_phasor_2016}, requiring microsecond-level synchronization accuracy. There are two popular approaches to achieve time synchronization.

The \textbf{network-based} approach achieves synchronization over a computer network by exchanging timing messages between the time server(s) and client devices. Where Ethernet connectivity is available at the meter location, and a time server is available in the same network, the network-based approach is suitable. The error introduced from network-based synchronization is dependent on a variety of factors: the protocol used, the time server jitter, network delay, and client device hardware. The accuracy of Network Time Protocol (NTP) is insufficient for phasor measurements. Precision Time Protocol (PTP) can achieve accuracy in the microsecond range (for client devices with software time-stamping) or nanosecond range (for devices with hardware time-stamping) due to its more sophisticated delay compensation methods. However, PTP requires an Ethernet Layer-2 network or multi-cast routing, which restricts where meters and time servers can be installed. Typically, this requires the meter and the time server to be located within the same building structure.

The \textbf{GPS-based} approach uses the satellite GPS signal to achieve synchronization. 
However, GPS signals are greatly diminished inside building structures, and unavailable in underground electrical rooms. Therefore, the GPS-based approach is best suited for outdoor installations. Notably, in network-based synchronization, time servers often use satellite GPS signals as the time source. GPS-based synchronization is generally highly accurate. Satellite atomic clocks are accurate beyond the nanosecond range and modern low-cost GPS receiver modules can typically achieve accuracies on the order of tens of nanoseconds \cite{berns2004gps}.

Our installation uses both of the above approaches. We installed GPS-disciplined PTP time servers located on the same Ethernet layer 2 network as the meters. The time servers receive GPS timing signals via satellite antennas and then exchange PTP timing messages with individual meters via the Ethernet network.

\subsection{Data Collection, Storage, and Hosting}

\begin{figure*}[ht!]
  \centering
  \includegraphics[width=1.\linewidth]{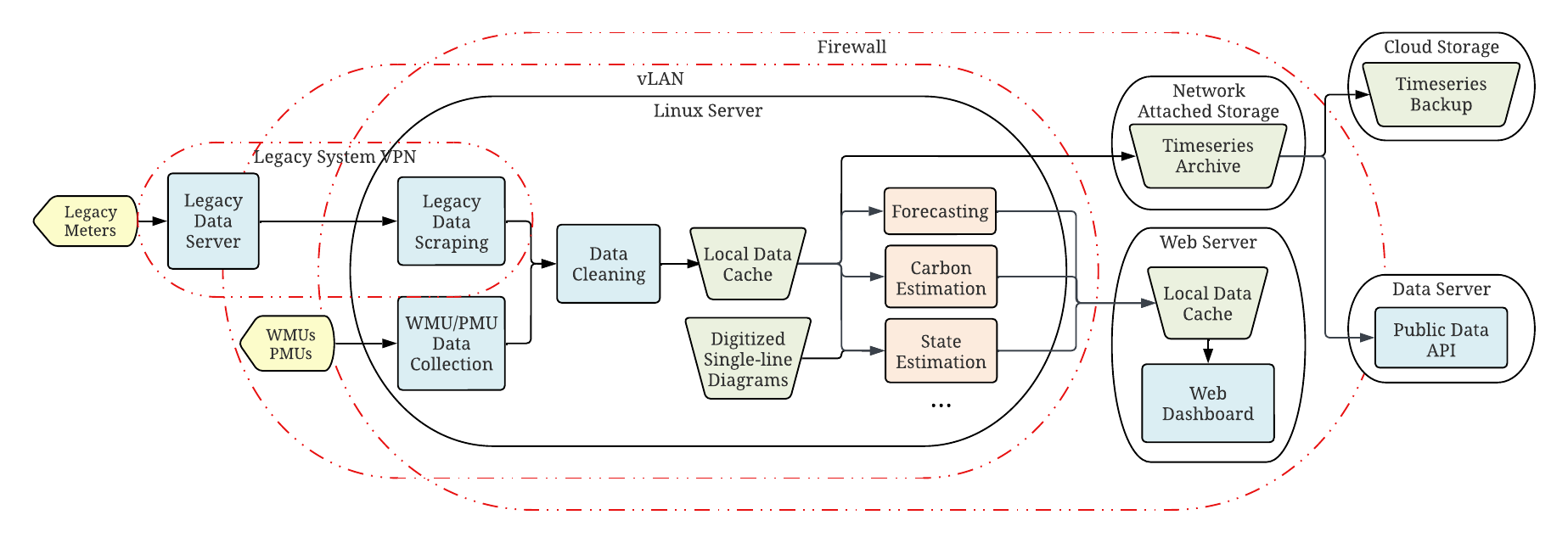}
  \caption{Cyber-physical infrastructure for data collection, processing, analysis, and hosting.}
  \label{fig:software_pipeline}
\end{figure*}

Depending on the availability of Ethernet network at the meter location, the communication to the meters is established via Ethernet or wireless cellular network. A central server collects data from all meters continuously. The entire computer network is protected by firewalls in a virtual local access network. Event-based data collection and processing is possible where edge computers are present at the meter location, which can reduce the required communication bandwidth. An illustration of the cyber-physical infrastructure is provided in \figref{fig:software_pipeline}. 

Due to the high sampling frequency, the collection and storage of data are non-trivial. For example, 20 sensors measuring 3-phase voltage and current waveforms at 2.5 kHz generate 0.3 million floating-point numbers each second. 
Topics such as event-based metering \cite{ge_power_2015}, compression \cite{wang_adaptive_2021, qiu_synchro-waveform_2024}, and specialized databases \cite{andersen_btrdb_2016} are beyond the scope of this paper.

\section{Accuracy and Error Characterization}
\label{sec:errors}
In this section, we discuss all known inaccuracies in this dataset. Compared to transmission systems, distribution systems generally require higher measurement accuracy due to short and lightly-loaded lines \cite{brouillon_bayesian_2021, stewart_phasor_2016}. We characterize measurement errors in our dataset by providing either an upper bound or error statistics. System topology and parameter inaccuracies are also discussed in this section.

\subsection{Sensors}
Sensor errors include the error from the meter instrument as well as current and potential transformers. Both phase and magnitude errors can be introduced. In our case, the meter instrument is certified to have a 0.5\% error upper bound (ANSI C12.20) \cite{egauge_systems_llc_accuracy_nodate}. For current and potential transformers, we document the specification in the digitized single-line diagrams where available. In general, the instrument transformers have errors bounded by 0.5-1\%. An exception is in lightly-loaded circuits, where over-sized current transformers are used to fit around the large physical conductors. The current may be below the detection limit. We make note of these cases in the released dataset. 

\subsection{Synchronization}
\label{sec:sync_err}

\begin{figure}[ht!]
  \centering
  \includegraphics{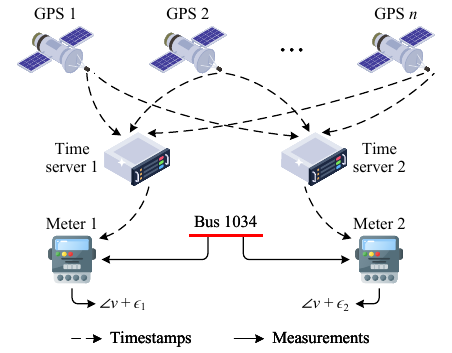}
  \caption{Synchronization test setup.}
  \label{fig:sync_test_setup}
\end{figure}

To measure the end-to-end synchronization error, we install pairs of two identical but independent network switches, time servers, and meters measuring the same bus voltage. The setup is illustrated in Figure \ref{fig:sync_test_setup}. 

\begin{figure}[t]
  \centering
  \includegraphics[width=0.85\linewidth]{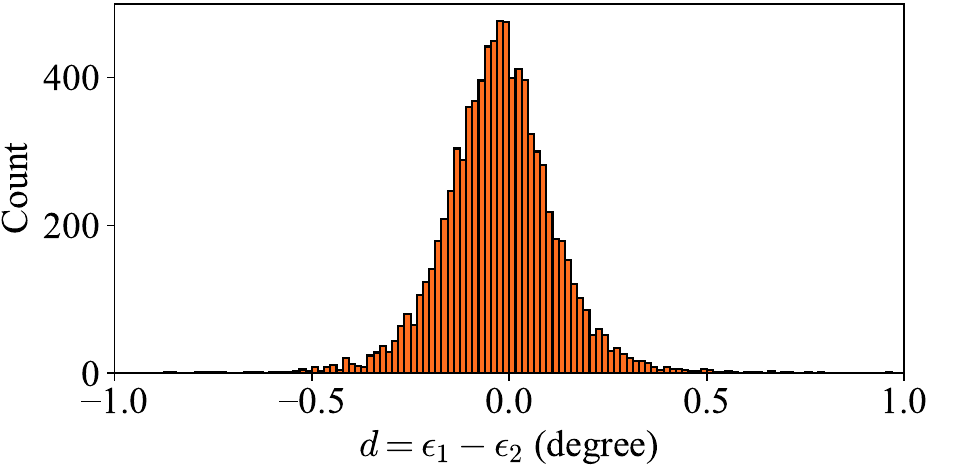}
  \caption{Synchronization error distribution.}
  \label{fig:sync_test_result}
\end{figure}

Denote the true voltage phase angle as $\angle v$ and the phase angle measurements from two meters as $\angle v + \epsilon_1$ and $\angle v + \epsilon_2$ at an instant in time. Assume the error terms
$\epsilon_1$ and $\epsilon_2$ are independent and identically distributed normal random variables with the same distribution $\mathcal{N}(0, \sigma^2)$. The phase angle difference between the two meters is therefore also normally distributed with variance $\sigma_d^2 = 2\sigma^2$:
\begin{subequations}    
\begin{align}
    \epsilon_1, \epsilon_2 &\sim \mathcal{N}(0, \sigma^2)\\
    d := \epsilon_1 - \epsilon_2 &\sim \mathcal{N}(0, \sigma_d^2).
\end{align}
\end{subequations}
The repeated measurements for $d$ are shown in Figure \ref{fig:sync_test_result}, from which $\sigma_d^2$ can be estimated empirically. Therefore, the distribution of $\epsilon_1$ and $\epsilon_2$ can be estimated even if the true voltage is unknown. 
In our experiment, 8610 samples were collected over 24 hours with $\sigma_d^2 \approx 0.221^\circ$ and thus $\sigma^2 \approx 0.110^\circ$.

\subsection{Topology and Parameters}
In practice, distribution system circuit models often contain errors due to out-dated or missing information. In the following, we describe where circuit information may be inaccurate.

For distribution lines, the conductor thickness and material are obtained from engineering drawings and are generally correct, but the insulation material and thickness are estimated from popular cable types given the voltage level. Line lengths are estimated as the Manhattan (taxicab) distance between the two terminals. The lengths of short lines (within the same building structure) are assumed to be zero. Lines are generally underground, un-transposed, and with unknown cable arrangement (e.g. on a cable tray).

For transformers, the nameplate ratings are generally accurate. The true transformer tap positions are unknown, although off-nominal tap positions are rare. Earth ground information is generally accurate, and grounding typically occurs on the Wye side of Delta-Wye transformers.

Other common issues in real-world systems include wrong current transformer ratio, polarity reversal, phase mis-labeling, and out-of-date switching status. This dataset is generally free of these issues as it has been validated using measurements, engineering drawings, and state estimation results.
\section{Applications}
\label{sec:applications}
Different applications require different system models. In this section, we formulate circuit models in phasor and time domains. We illustrate the usefulness of this dataset via state estimation in both phasor and time domains, since state estimation requires accurate circuit parameters, dense sensor deployment, and synchronization. In particular, the state estimation algorithms as formulated here are not possible with existing SCADA, AMI and BMS systems due to the lack of phase angles in phasor domain and the lack of synchronization in time domain.

The set of all possible model formulations and potential applications extends far beyond the two shown here. A list of relevant applications is provided in Table \ref{tab:applications}. For additional discussion on PMU applications, see recent surveys such as \cite{hojabri_comprehensive_2019, dusabimana_survey_2020} and the references therein. As the possible use cases are broad, it is impractical to exhaustively discuss and implement all potential test cases and applications. We refer interested users to the code examples provided as a starting point for developing a particular application.

The distribution system can be abstracted as a graph $G=(N,E)$, in which nodes $N$ represent buses, and edges $E$ represent power transfer elements such as lines, transformers, and switches. Let $n = |N|$ be the total number of nodes, and $m = |E|$ be the total number of edges. The graph is directed with $(j,k)\in E$ indicate direction $j\rightarrow k$. Edge directions can be arbitrarily assigned. In the following, we first develop component models for each edge $(j,k)\in E$, and then combine them to form a network model.

\subsection{Phasor Domain State Estimation}
\subsubsection{Phasor Domain Circuit Model}
\label{sec:model_phasor}
\begin{figure}[h!]
    \centering
    \includegraphics{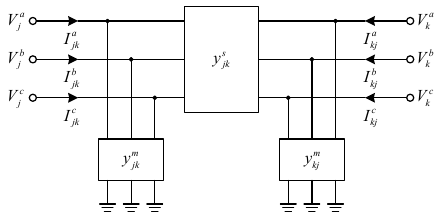}
    \caption{Distribution line model (3-phase, 3-wire).}
    \label{fig:phasor_line_models}
\end{figure}

Here, we consider a multi-phase\footnote{The examples are illustrated with a 3-phase 3-wire circuit. However, the approach is generally applicable to arbitrary multi- and single-phase components such as 3-phase 4-wire lines and split-phase transformers.} 
unbalanced circuit consisting of lines and transformers. 
A \textbf{multi-phase line} connecting buses $j$ and $k$ can be modeled by a $\Pi$ circuit (\figref{fig:phasor_line_models}). The line parameters are series admittance $y^s_{jk} = y^s_{kj} \in \mathbb{C}^{3\times 3}$ and shunt admittances $y^m_{jk}, y^m_{kj} \in \mathbb{C}^{3\times 3}$. The current injections $I_{jk}, I_{kj} \in \mathbb{C}^3$ and the bus voltages $V_j, V_k \in \mathbb{C}^3$ are linearly related by an admittance matrix $\bm{Y}_{jk} \in \mathbb{C}^{6\times 6}$.
\begin{align}
    \begin{bmatrix}I_{jk}\\I_{kj}\end{bmatrix}
     = 
    \begin{bmatrix}
        y_{jk}^s + y_{jk}^m & -y_{jk}^s\\
        -y_{kj}^s & y_{kj}^s + y_{kj}^m
    \end{bmatrix}
    \begin{bmatrix}V_j\\V_k\end{bmatrix}
     =: 
     \bm{Y}_{jk} \begin{bmatrix}V_j\\V_k\end{bmatrix}
\end{align}

\begin{figure}[h!]
    \centering
    \includegraphics{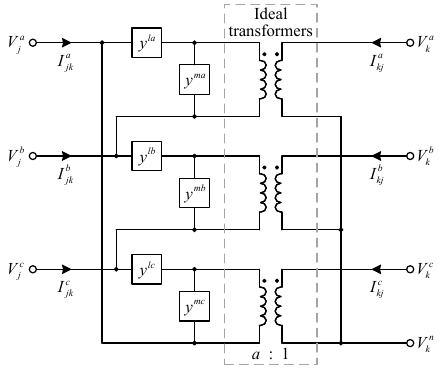}
    \caption{Delta-Wye transformer model.}
    \label{fig:phasor_xformer_models}
\end{figure}

The unitary voltage network offers a flexible method for modeling \textbf{multi-phase transformers} including 3-phase, single-phase, and split-phase arrangements \cite{coppo_generalised_2017, moorthy_new_2002}. 
For example, 3-phase Delta-Wye transformers (\figref{fig:phasor_xformer_models}) can be modeled by an admittance matrix $\bm{Y}_{jk} \in \mathbb{C}^{6\times 6}$. Given turns ratio $a\in\mathbb{R}$, series admittance $y^l \in\mathbb{C}^{3\times3}$, and shunt admittance $y^m \in\mathbb{C}^{3\times3}$, the nodal variables have the relationship:
\begin{subequations}
\begin{align}
    \label{eqn:dy_xfmr}
    \begin{bmatrix}I_{jk}\\I_{kj}\end{bmatrix}
     &= 
    \begin{bmatrix}
        \Gamma^\top y^l \Gamma & -\Gamma^\top ay^l\\
        -ay^l\Gamma & a^2(y^l + y^m)
    \end{bmatrix}
    \begin{bmatrix}V_j\\V_k\end{bmatrix}
    =: 
    \bm{Y}_{jk} 
    \begin{bmatrix}V_j\\V_k\end{bmatrix}
    \\
    \Gamma &= \begin{bmatrix}
        1 & -1 & 0 \\
        0 & 1 & -1 \\
        -1 & 0 & 1 
    \end{bmatrix}.
\end{align}
\end{subequations}

The line and transformer models above can each be uniquely characterized by 4 parameters: $(y_{jk}^s, y_{kj}^s, y_{jk}^m, y_{kj}^m)$, which are the series admittance in the $j\rightarrow k$ direction, the series admittance in the $k\rightarrow j$ direction, the sending end shunt admittance, and the receiving end shunt admittance. These 4 parameters offer a unified description of branch components in a graph. For lines, the parameters have a physical interpretation as they correspond to the admittances in the $\Pi$ circuit model. For transformers, the parameters are the result of the following derivation, and there is no direct physical interpretation. Rearranging $\bm{Y}_{jk}$ in \eqref{eqn:dy_xfmr}, we have
\begin{subequations}
\begin{align}
    \bm{Y}_{jk}
    & = 
    \begin{bmatrix}
        y_{jk}^s + y_{jk}^m & -y_{jk}^s\\
        -y_{kj}^s & y_{kj}^s + y_{kj}^m
    \end{bmatrix}\\
    y_{jk}^s & :=\Gamma^{\sf T} a y^l, \\
    y_{kj}^s &:=a y^l \, \Gamma,  \\
    y_{jk}^m &:=\Gamma^{\sf T} y^l ( \Gamma - a), \\
    y_{kj}^m &:=a y^l ( a - \Gamma ) + a^2 y^m.
\end{align}
\end{subequations}

Given the 4 parameters for each edge component, we now derive the overall network admittance matrix $\bm{Y} \in \mathbb{C}^{3n\times 3n}$. Denote $[\bm{Y}]_{j,k}\in\mathbb{C}^{3\times 3}$ as the matrix at the $j$-th row block and $k$-th column block of $\bm{Y}$. Each block matrix is constructed as such:
\begin{align*}
[\bm{Y}]_{j, k} & \ := \ \left\{ \begin{array}{lcl}
			- y^s_{jk},  & & j\sim k \ \ (j\neq k)  \\
			\sum_{l: j\sim l}\left( y^s_{jl} +  y_{jl}^m \right),   & & j=k   \\
			0   & & \text{otherwise}
			\end{array}  \right.
\end{align*}

Denote the vector of 3-phase nodal voltages as $\bm{V}\in \mathbb{C}^{3n}$ and the vector of 3-phase nodal current injections as $\bm{I}\in \mathbb{C}^{3n}$. Then, $\bm{I}$, $\bm{V}$ are related via
\begin{align}
    \label{eqn:bim}
    \bm{I} &= \bm{Y} \bm{V}.
\end{align}
The bus injection model in the phasor domain is useful for steady-state analysis where the voltage and current waveforms are well-approximated by phasors. In addition, if the variables of interest are currents and voltages, the $I$-$V$ model is linear. 


\subsubsection{Phasor Domain State Estimation}
\label{sec:state_estimation}

Assume the voltage and current injection phasors are measured for a subset of nodes shown in red in \figref{fig:data_topology}. The problem is to recover the unknown voltage and current injection phasors for all nodes, during steady-state operation. 
The set of nodes where currents are measured and those where voltages are measured may be different.
Let $\bm{I}_{1}, \bm{I}_{2}$ be the measured and unmeasured nodal current injections and 
$\bm{V}_{1}, \bm{V}_{2}$ the measured and unmeasured nodal voltages. Then there exist permutation matrices $\bm{P}_I$ and $\bm{P}_V$ such that 
\begin{align}
    \begin{bmatrix}\bm{I}_1\\ \bm{I}_2\end{bmatrix} = \bm{P}_I\bm{I}, \qquad
    \begin{bmatrix}\bm{V}_1\\ \bm{V}_2\end{bmatrix} = \bm{P}_V\bm{V}.    
\end{align}
Substituting this into the system model \eqref{eqn:bim} yields
\begin{align}
    \label{eqn:bim_partition}
    \begin{bmatrix} \bm{I}_{1} \\ \bm{I}_{2}\end{bmatrix} 
    &= \begin{bmatrix} \bm{Y}_{11} & \bm{Y}_{12} \\ \bm{Y}_{21} & \bm{Y}_{22}\end{bmatrix} 
    \begin{bmatrix} \bm{V}_{1} \\ \bm{V}_{2}\end{bmatrix} 
\end{align}
where the matrix in \eqref{eqn:bim_partition} is $\bm{P}_I\bm{Y}\bm{P}_V^{-1}$. 

Given measurement $\bm{I}_{1}, \bm{V}_{1}$ and network admittance $\bm{Y}$, we would like to recover the unknown quantities $\bm{I}_{2}, \bm{V}_{2}$. Choose nodal voltage estimation $\hat{\bm{V}}$ as the system state, then the state estimation can be formulated as a least-squares problem:
\begin{align}
    \begin{split}
    & \argmin_{\hat{\bm{V}}_{1}, \hat{\bm{V}}_{2}}
    {\norm{
        \begin{bmatrix} \bm{V}_{1} \\ \bm{I}_{1}\end{bmatrix} 
        - 
        \begin{bmatrix} \mathbb{I} & \bm{0} \\ \bm{Y}_{11} & \bm{Y}_{12} \end{bmatrix}
        \begin{bmatrix} \hat{\bm{V}}_{1} \\ \hat{\bm{V}}_{2}\end{bmatrix}
    }_2}
    \end{split}
\end{align}
where $\hat{\bm{V}}_{1}, \hat{\bm{V}}_{2}$ are the nodal voltages for nodes with and without voltage measurements, and $\mathbb{I}$ is the identity matrix of the appropriate size. The measurement Jacobian matrix is full-rank for this dataset.

\begin{figure}[t]
    \centering
    \includegraphics[width=1.0\linewidth]{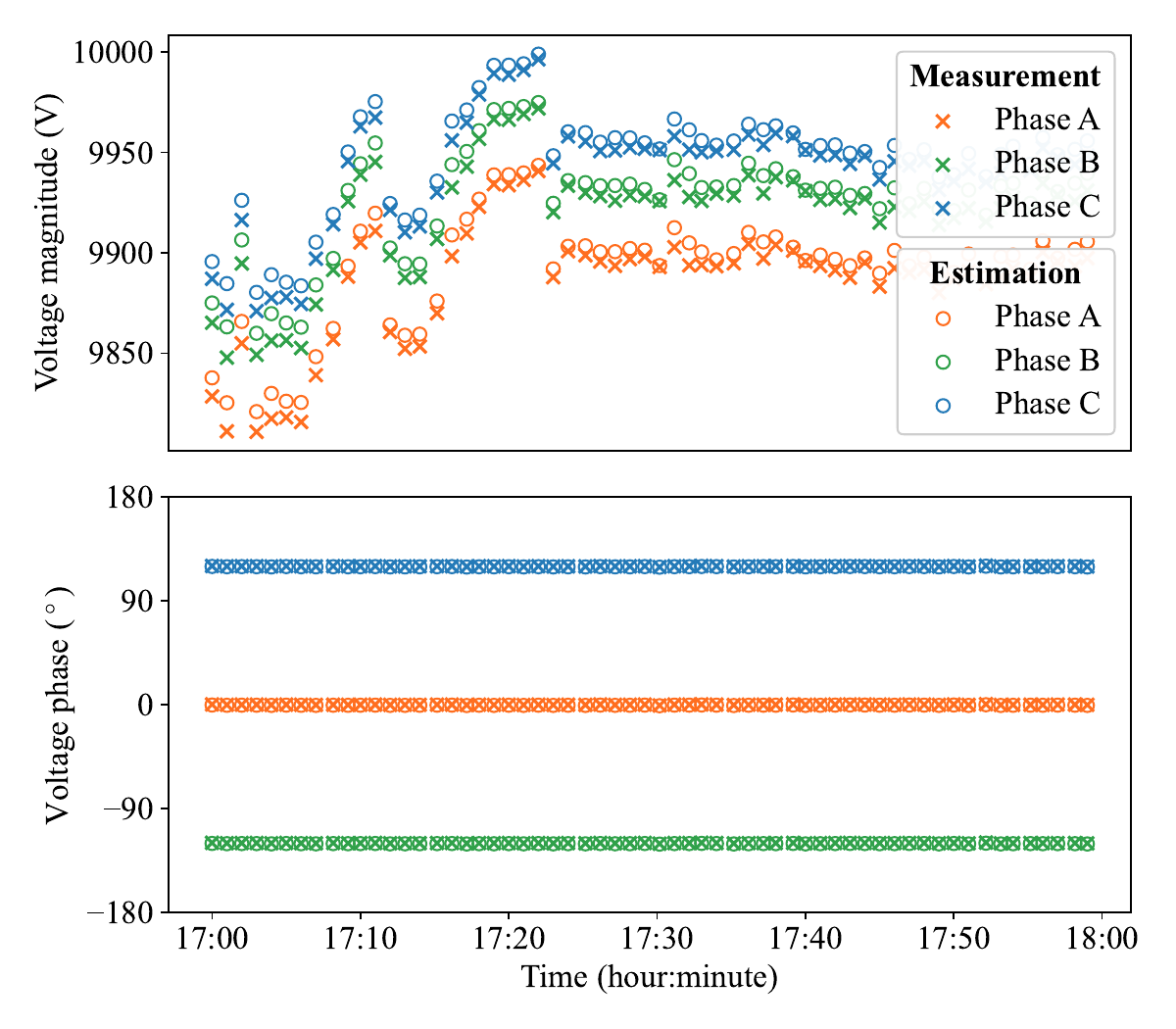}
    \caption{Bus 1033 voltage phasor.}
    \label{fig:state_est_results}
\end{figure}

The least-squares problem may be additionally weighted by measurement accuracy and measured magnitudes. We show the state estimation result on the A-side circuit in \figref{fig:state_est_results}. The average normalized residual error (objective value) of all the measured quantities on 228 time points collected over an hour is $0.851\%$.

\subsection{Time Domain State Estimation}
\subsubsection{Time Domain Circuit Model}
\label{sec:model_time_domain}

\begin{figure}[h!]
    \centering
    \includegraphics{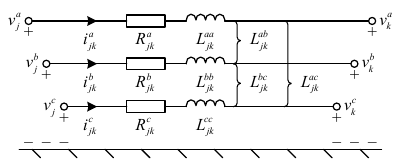}
    \caption{Three-phase (inductive) line model.}
    \label{fig:time_domain_line_model}
\end{figure}

Following \cite{pfeifer_time-domain_2021}, we consider a circuit of multi-phase distribution lines (\figref{fig:time_domain_line_model}). Each line connecting buses $j$ and $k$ is modeled by series resistances $\bm{R}_{jk} \in \mathbb{R}^{3\times 3}$ and inductances $\bm{L}_{jk} \in \mathbb{R}^{3\times 3}$, where
\renewcommand{\arraystretch}{1.4}
\begin{subequations}
\begin{align}
\bm{R}_{jk} &= 
\begin{bmatrix}
R_{jk}^{a} &0 &0\\
0 &R_{jk}^{b} &0\\
0 &0 &R_{jk}^{c}
\end{bmatrix}
\\
\bm{L}_{jk} &= 
\begin{bmatrix}
L_{jk}^{aa} &L_{jk}^{ab} &L_{jk}^{ac}\\
L_{jk}^{ab} &L_{jk}^{bb} &L_{jk}^{bc}\\
L_{jk}^{ac} &L_{jk}^{bc} &L_{jk}^{cc}
\end{bmatrix}.
\end{align}
\end{subequations}

Now consider a network of such lines. Denote the node-by-line incidence matrix as $\bm{C}\in\{0, \pm 1\}^{n\times m}$. The relationship between line voltage drop $\bm{v}_{l}\in\mathbb{R}^{3m}$ and nodal voltages $\bm{v}_{b}\in\mathbb{R}^{3n}$ obey Kirchhoff's voltage law \eqref{eqn:dse_KVL}. Similarly, line currents $\bm{i}_{l}\in\mathbb{R}^{3m}$ and nodal current injections $\bm{i}_{b}\in\mathbb{R}^{3n}$ obey Kirchhoff's current law \eqref{eqn:dse_KCL}:\footnote{Here we assumed for simplicity that the current flow from the two ends of a line are equal in magnitude. For lines with unequal shunt elements and transformers, \eqref{eqn:dse_KCL} can be extended by keeping track of the current flow from both terminals of each line.}

\begin{subequations}
\begin{align}    
\label{eqn:dse_KVL}
\bm{v}_{l} &= \left( \bm{C} ^\mathsf{T} \otimes \mathbb{I}_3 \right) \bm{v}_{b}
\\
\label{eqn:dse_KCL}
\bm{i}_{b} &= \left( \bm{C} \otimes \mathbb{I}_3 \right) \bm{i}_{l}
\end{align}
Here
\label{eqn:dse_KCLKVL}
\end{subequations}
 $\mathbb{I}_3$ is the $3\times3$ identity matrix and $\otimes$ denotes the Kronecker product. 

Finally, the voltage drop across lines follows Ohm's law
\begin{subequations}
    \label{eqn:dse_ohms}
    \begin{gather}
    \bm{v}_{l} = \bm{R}\bm{i}_{l} + \bm{L}\dot{\bm{i}}_{l}
    \\
    \bm{R}=blockdiag(\bm{R}_{jk} : (j, k)\in E)
    \\
    \bm{L}=blockdiag(\bm{L}_{jk} : (j, k)\in E)
    \end{gather}
\end{subequations}

Combine \eqref{eqn:dse_KVL}, \eqref{eqn:dse_ohms} and let $\hat{\bm{C}} = \bm{C} ^\mathsf{T} \otimes \mathbb{I}_3$, then we have the network dynamics in terms of bus voltages $\bm{v}_{b}$ and line currents $\bm{i}_{l}$:
\begin{equation}
\label{eqn:line_dynamics}
\dot{\bm{i}}_{l} = -\bm{L}^{-1}\bm{R}\bm{i}_{l} + \bm{L}^{-1}\hat{\bm{C}}\bm{v}_{b}
\end{equation}
Together, \eqref{eqn:dse_KCL} and \eqref{eqn:line_dynamics} (or equivalently \eqref{eqn:dse_KCLKVL} and \eqref{eqn:dse_ohms}) provide a complete characterization of the circuit. Notice here the circuit model is defined for arbitrary (not necessarily sinusoidal) time domain signals $\bm{v}_{b}(t), \bm{v}_{l}(t), \bm{i}_{b}(t), \bm{i}_{l}(t)$. 

\subsubsection{Time domain State Estimation}
\begin{figure}[t]
    \centering
    \includegraphics[width=1.0\linewidth]{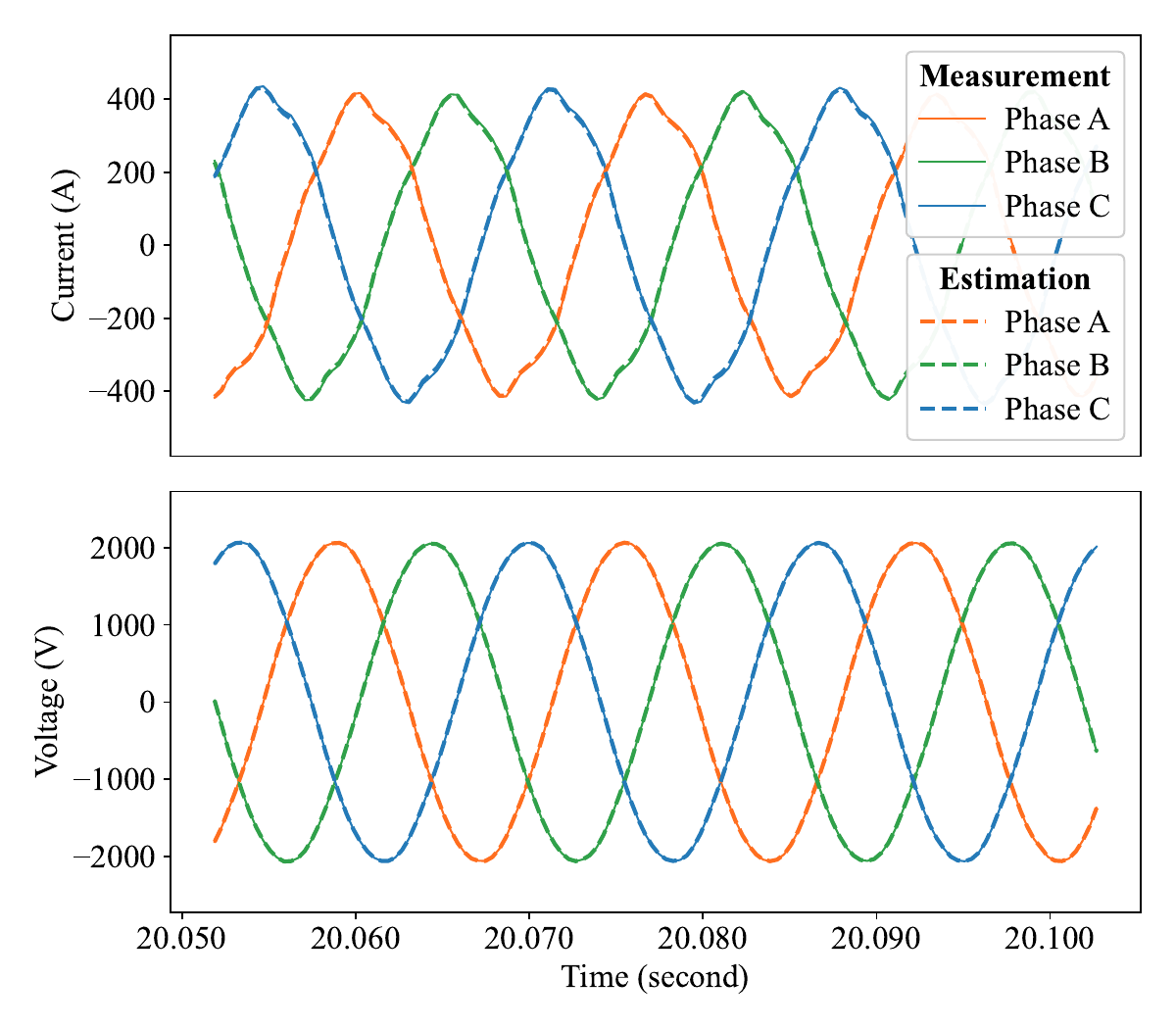}
    \caption{Time domain voltage and current injection at bus 1034.}
    \label{fig:dse_result}
\end{figure}

State estimation in time domain enables the reconstruction of sub-cycle events and instantaneous power flow for un-observed current and voltage signals. Given voltage and current waveforms on a subset of the network, we recover the nodal voltage and branch current for all network elements. In this example, we again formulate a least squares optimization problem, where the (discrete-time version of the) linear circuit model in \secref{sec:model_time_domain} is used as the constraint:
\begin{equation}
\begin{array}{cl}
\min\limits_{\hat{\bm{i}}_{l} [k], \hat{\bm{i}}_{b} [k], \hat{\bm{v}}_{b} } 
& \displaystyle\sum_{k=1}^{T} \left\| 
\begin{bmatrix}\bm{i}_l[k]\\\bm{i}_b[k]\\\bm{v}_b[k]\\\end{bmatrix} - 
\begin{bmatrix}\bm{H}_1\\\bm{H}_2\\\bm{H}_3\\\end{bmatrix}
\begin{bmatrix}
\hat{\bm{i}}_{l} [k] \\
\hat{\bm{i}}_{b} [k] \\
\hat{\bm{v}}_{b} [k]
\end{bmatrix} \right\|_{2}\\
\text{s.t.} & \hat{\bm{i}}_{l} [k+1] = \bm{A}_{d}\hat{\bm{i}}_{l} [k] + \bm{B}_{d}\hat{\bm{v}}_{b} [k]\\
& \hat{\bm{i}}_{b} [k] = \hat{\bm{C}}^\mathsf{T} \hat{\bm{i}}_{l} [k]
\end{array}
\end{equation}
where 
$\bm{H}_1, \bm{H}_2, \bm{H}_3$ are matrices whose rows are standard Euclidean basis vectors that select the variables which are measured. $\bm{i}_{l}, \bm{i}_{b}, \bm{v}_{b}$ are measured values, $\bm{A}_{d}$ and $\bm{B}_{d}$ are obtained by discretizing the time domain model in \eqref{eqn:line_dynamics} via zero-order hold. The results following this formulation are shown in \figref{fig:dse_result}. The average normalized residual error (objective value) of all measured quantities across 864,000 time samples collected over an hour is $1.23\%$.

\subsection{Adaptive Predictive Voltage Control}
Here we briefly describe a case study using this dataset for distribution grid voltage control \cite{cui2025leveraging}. We see that the dataset plays a key role in each step of analysis, modeling, planning, and operations.

\paragraph{Analysis}
The dense phasor measurement timeseries indicate that distribution-level loads are highly time-varying due to PV generation and EV charging. The dense spatial coverage of the dataset reveals that traditional voltage control algorithms designed under the constant-load assumption can exhibit instability when deployed in such environments \cite{cui2025leveraging}.

\paragraph{Modeling}
The digitized circuit topology and parameters provided in the dataset are used to compute the network resistance and reactance matrices \(R\) and \(X\) in the Linear DistFlow model. Moreover, the abundance of offline historical time-series data is essential for training accurate prediction models for future load and generation, which are explicitly incorporated in the controller design \cite{cui2025leveraging}.

\paragraph{Planning}
For decentralized voltage controller design, control parameters must be tuned offline to ensure robustness across a wide range of operating conditions. The historical data in this dataset enables systematic tuning and evaluation of controller performance under realistic variability in PV output, EV demand, and topology changes.

\paragraph{Operations}
Finally, once deployed, real-time phasor data are required closed-loop inverter-based voltage control. \cite{cui2025leveraging} presents a detailed case study using this dataset, including numerical simulation results that validate controller performance under real-world operating conditions. Together, these steps illustrate a complete life cycle of problem identification, modeling, planning, and deployment in a distribution system, using this dataset.

\section{Conclusion}
\label{sec:conclusion}
This work introduces the SoCal 28-Bus Dataset, an open-access dataset comprising synchronized phasor and waveform measurements from a real-world electrical distribution network. The dataset is distinguished by its high spatial and temporal resolution, time synchronization, and the inclusion of detailed circuit topology and parameters. We demonstrate its utility through phasor- and time-domain state estimation. By bridging the gap between simulated test cases and real-world measurement data, we hope this dataset will accelerate the development and validation of algorithms for monitoring, control, and planning in modern distribution grids.

\section*{Acknowledgments}
We thank our partner distribution system operators for their generous in-kind contribution during hardware installation. We thank George Lee for his technical guidance. We thank Nicolas Christianson and Verena Häberle for their comments and critiques during manuscript preparation.

\appendices

\begin{table*}[b]
\centering
\caption{Materials and Labor Cost Estimation (per Meter Installation)\label{tab:material_cost}}
\begin{tabular}{|p{5cm}|p{1.8cm}|p{0.8cm}|p{6.2cm}|}
    \hline
    Item & Price & Qty & Comment\\\hline
    Meter & \$750 - 1250
    & 1 & EGauge EG4X used in this project. Alternatives exist.\\\hline
    PTP time server & \$550 - 3000 & 1 & Can serve multiple meters. Raspberry Pi CM4 may also be configured as a GPS-enabled PTP time server.\\\hline
    Current transformer & \$20 - 300 & 3 - 4 & More needed for sub-metering.\\\hline
    Potential transformer & varies & 0 - 1 & Not required for low-voltage metering. Often already exists for high voltage switchgear. \\\hline
    Waterproof enclosure & \$20 - 100 & 1 & \\\hline
    Conduits & \$1 / foot & varies & For Ethernet cable and/or CT leads.\\\hline
    Ethernet cable & \$0.1 - 0.2 / foot & varies & \\\hline
    Cellular data module & \$20 - 200 & 0 - 1 & Required for remote locations with no Ethernet connectivity. Data transmission cost can be significant.\\\hline
    Miscellaneous hardware (breakers, wires, terminal blocks, DIN rails, etc.) & varies & varies & Cost is low when bought in bulk. \\\hline
    Edge computing equipment & \$20 - 200 & 0 - 1 & Optional. e.g. Raspberry Pi.\\\hline
    Labor (electricians and project manager) & \$50 - 100 / hour & varies & Depends on installation complexity. Can be significant.\\\hline
\end{tabular}
\end{table*}

\begin{table*}[b]
\centering
\caption{Relevant Applications\label{tab:applications}}
\begin{tabular}{m{0.2cm}|p{3.5cm} p{4cm} p{4cm} p{3.5cm}}
 & Monitoring & Diagnostics & Control & Planning \\
\hline
\multirow{7}{*}[-8pt]{\rotatebox{90}{Phasor}} & State estimation & Event detection \& localization & Voltage control & Preventative maintenance \\
 & Constraint violation & System identification & Optimal dispatch & Capacity expansion \\
 & Power quality & Outage detection & Network reconfiguration & Hosting capacity\\
 & Asset loading/overload & Phase identification & Islanding & Non-wire alternatives \\
 & Energy theft detection & Island detection & Virtual power plants & \\
 & Data-driven prediction & Model validation & & \\
 & Billing & Fault ride-through & & \\
 & Carbon calculation & & & \\
\hline
\multirow{6}{*}[+2pt]{\rotatebox{90}{Waveform}} & Incipient fault detection & Transient event analysis & Frequency control & \\
 & Power quality & IBR dynamics modeling & Dynamic virtual power plants & \\
 & State estimation & Oscillation detection & & \\
 & Frequency estimation & System identification & & \\
 & Inertia \& stability & & & \\
\end{tabular}
\end{table*}
\section*{Materials and cost}
For power utilities, microgrid operators, and researchers interested in deploying similar metering infrastructure, we welcome discussions and are open to provide technical consultation. We provide an itemized cost estimation in Table \ref{tab:material_cost}. 
The cost will vary significantly depending on local material and labor cost, and site-specific complexities. Important considerations are: indoor vs. outdoor, the voltage level, availability of Ethernet, and whether de-energization is required.

\section*{Dataset Applications}
In Table \ref{tab:applications}, we provide a non-exhaustive list of applications for distribution system synchrophasor and synchrowaveforms. These applications may also be used by power utilities and microgrids to evaluate the economic benefit and justify the cost of deploying such a sensor network in their system. 


\bibliographystyle{IEEEtran}
\bibliography{BibliographyNew}


 
\vspace{11pt}



\vfill

\end{document}